\documentclass[11pt]{article}
\usepackage{amsthm}
\usepackage{amsmath,amssymb}
\title{\large \bf Algorithms for enumerating and counting \\ D$2$CS of some graphs}
\author{\small
P.Venkata Subba Reddy and K.Viswanathan Iyer \thanks{author for correspondence}\\ 
\small Department of Computer Science and Engineering\\  
\small National Institute of Technology\\  
\small Tiruchirapalli 620 015, India.\\
\small email : venkatpalagiri@gmail.com, kvi@nitt.edu}    
\date{}  
\begin{document}
\maketitle
\begin{abstract} 
We define a D$2$CS of a graph $G$ to be a set $S \subseteq V(G)$ with diam$(G[S]) \leq 2$. A D$2$CS arises in connection with conditional coloring and radio-$k$-coloring of graphs. We study the problem of counting and enumerating D$2$CS of a graph. We first prove the following propositions:\\
(1) Let $f(k,h)$ be the number of D$2$CS of a complete $k$-ary tree of height $h$. Then \\
\hspace*{1cm}$f(k,h)=\frac{k}{k-1}(f(k+1,1)-4)(k^{h-1}-1)+f(k,1)$ and $f(k,1)=2^k+k+1.$\\
(2) A  Fibonacci tree, a variant of a binary tree is defined recursively as follows: 
(a)  Fibonacci tree of order $0$ and $1$ is a single node.
(b)  Fibonacci tree of order $n$ ($n \geq 2$) is constructed by attaching tree of order $n-2$ as the leftmost child of the tree of order $n-1$.\\
Let $g(n)$ denote the number of D$2$CS in a  Fibonacci tree of order $n$. Then \\
\hspace*{4cm}$g(n)= 3 \cdot 2^{n-2}-(F_{n-1}+F_{n+1})+2.$ \\
(3) A binary Fibonacci tree of order $n$ ($n >1$) is a variant of a binary tree whose left subtree is of order $n-1$ and right subtree of order $n-2$. An order $0$ Fibonacci tree has a single node, and  an order $1$ tree is $P_2$. \\
Let $h(n)$ denote the number of D$2$CS in a binary Fibonacci tree of order $n$. Then\\ 
\hspace*{4cm} $h(n)=2F_n+3F_{n+2}-9.$ \\
(4) A binomial tree $B_k$ of order $k$ ($k \geq 0)$ is an ordered tree defined recursively as: 
(i)  $B_0$ is a one-vertex graph.
(ii) $B_k$ consists of two copies of $B_{k-1}$ such that the root of one is the leftmost child of the root of the other.\\
Let $b(k)$ denote the number of D$2$CS in a  binomial tree $B_k$. Then  
 $b(k)=k2^k+2.$  \\
(5) Let $G$ be a split graph with $K \subseteq V(G),|K|=\omega(G)=k$, for all $v \in K$ and $r > 1$, $d(v)=k+r-1$ and for all $v' \in V(G) \setminus K, d(v')=1$. Then   the number of D$2$CS in $G$ is $k2^{k-1}(2^r-1)+2^k+kr$. \\
We then show: all D$2$CS in a given graph $G$ with $n$ vertices can be enumerated in time O$( n^3/log ^ 2 n)$ for each D$2$CS. 
We finally show: all maximal D$2$CS in a strongly chordal graph on $n$ vertices can be enumerated and counted in time O($n$).

\end{abstract}
\section{Introduction} 
Let $G= (V(G),E(G))$ be a simple, connected, undirected graph. For a vertex $v \in V(G)$, the (\textit{open}) \textit{neighborhood} of $v$ in $G$ is $N(v)$= \{$u \in V(G):(u,v) \in E(G)$\}, the \textit{closed neighborhood} of $v$ is defined as $N[v]=N(v) \cup \{v\}$ and the degree of $v$ is $d(v)=|N(v)|$. The subgraph induced by a set $A \subseteq V(G)$ is denoted by $G[A]$. Let $\omega(G)$ (or simply $\omega$) denote the clique number of a graph $G$. The \textit{square} of a graph $G$, denoted by $G^2$, has the same vertex set as $G$, and the edge set $E(G^2)=\{(u,v): d(u,v) \leq 2 \; \text{in} \; G\}$. The distance $d(u,v)$ between two vertices $u$ and $v$ is the minimum length of a path between $u$ and $v$. The \textit{diameter} of a graph $G$ is $diam(G)$=max \{$d(u,v):u,v \in V(G) $\}. For a vertex $v \in V(G)$ we define by $N^k(v)=$ \{$u :u \in N(v), u > k $\}. We define a \textit{distance}-$2$-\textit{clique-set} (D$2$CS) of a graph $G$ as a subset $S$ of $V(G)$ such that every two distinct vertices in $S$ are at a distance at most two in $G[S]$ i.e., $diam(G[S]) \leq 2$. For example $K_{1,n}$ has $2^n+n+1$ D$2$CS.  A D$2$CS is \textit{maximal} if it is not properly contained in any other D$2$CS. A \textit{maximum} D$2$CS is one which  has the largest size among all D$2$CS. For undefined terms and notations see standard texts in graph theory such as ~\cite{Bondy,Gol}. \\
Recently counting and enumeration of certain specified sets in a graph have been widely investigated e.g., in data mining. In this paper we deal with the problem of conting and enumeration of D$2$CS of general and some restricted class of graphs. Unlike a clique and an independent set every subset of a D$2$CS need not be a D$2$CS. In general, the problem of finding all D$2$CS is difficult. A graph can have an exponential number of D$2$CS. For example the complete graph $K_n$ on $n$ vertices has $2^n$, a $k$-tree on $n$ vertices has $2^n(1-2^{-(k+1)})+n-k$, the ladder graph $L_n \cong P_n \Box P_2$ has $10n-6$ and the graph $\overline {K_n}$ of $n$ vertices has just $n+1$ D$2$CS. The number of  D$2$CS in any other graph with $n$ vertices lies between O$(n)$ and O$(2^n)$.
\section{D$2$CS of some structured graphs} 
We first give our results for the cases when the graph $G$ is a : complete $k$-ary tree, Fibonacci tree, binary Fibonacci tree and binomial tree.
For two integers $k,h > 0$, let $f(k,h)$ be the number of D$2$CS of a complete $k$-ary tree of height $h$. Then we have 
\begin{align*}
f(k,h)&=f(k,h-1)+k^{h-1}(2^{k+1}+k-2), \quad h > 1, \text{and} \\    
f(k,1)&=2^k+k+1. 
\end{align*}
\noindent By solving the above recurrence we get 
\begin{equation}
f(k,h)=\frac{k}{k-1}(f(k+1,1)-4)(k^{h-1}-1)+f(k,1).
\end{equation}
\newtheorem{pro1}{Proposition}  
\begin{pro1}
Let $f'(k,h)$ be the number of D$2$CS of a rooted tree $T$ with $\Delta(T)=k$ and height $h$. Then $2^k+k+3h-5 \leq f'(k,h) \leq (2^k+k-3)(2+l)+4$, where $l=\frac{1}{k-2}(k(k-1)^{h-2}-2)(k-1)$.
\end{pro1}
\begin{proof}
It can be easily checked that $f'(k,h) \geq  2^k+k+3h-5$. By $f_{max}'(k,h)$ we denote the maximum possible number of D$2$CS of a tree $T$ with $\Delta(T)=k$ and height $h$. Considering the root at level $0$, a tree $T$ with $\Delta(T)=k$ and height $h$ has maximum number of D$2$CS only if it has $k(k-1)^{i-1}$ vertices at level $i$, for all $1 \leq i \leq h$. Then it is easy to see that 
$$f_{max}'(k,h)=f(k-1,h)+f(k-1,h-1)+2^k-2.$$
\noindent By virtue of Eq.(1) we obtain $f_{max}'=(2^k+k-3)(2+l)+4$, where $l=\frac {k-1}{k-2}(k(k-1)^{h-2}-2)$.
\end{proof} 

\subsection{Fibonacci Trees and Binomial Trees}
 \textbf{Definition 1.} A  Fibonacci tree, a variant of a binary tree is defined recursively as follows: 
(a)  Fibonacci tree of order $0$ and $1$ is a single node.\\
(b)  Fibonacci tree of order $n$ ($n \geq 2$) is constructed by attaching tree of order $n-2$ as the leftmost child of the tree of order $n-1$.\\
Let $g(n)$ denote the number of D$2$CS in a  Fibonacci tree of order $n$. Then,
$$g(n)=g(n-1)+g(n-2)+3*2^{n-4}-2, \quad n \geq 4,  $$ 
with the initial conditions $g(2)=2$ and $g(3)=4$.\\
The (ordinary) generating function $G(z)$ for the sequence $g(n)$ is given by 
$$G(z)= \frac {5z^3-z^2-4z+2}{(2z^2-3z+1)(1-z-z^2)}.$$ 
It then follows that
$$g(n)= 3*2^{n-2}-(F_{n-1}+F_{n+1})+2=3*2^{n-2}-L_n+2,$$
where $L_{n}$ is the $n^{\rm th}$ Lucas number.\\
 \textbf{Definition 2.}  A binary Fibonacci tree of order $n$ ($n >1$) is a variant of a binary tree whose left subtree is of order $n-1$ and right subtree of order $n-2$. An order $0$ Fibonacci tree has a single node, and  an order $1$ tree is $P_2$. \\
Let $h(n)$ denote the number of D$2$CS in a binary Fibonacci tree of order $n$. Then 
$$h(n)=h(n-1)+h(n-2)+9, \quad n \geq 5  $$ 
with the initial conditions $h(3)=10$ and $h(4)=21$.\\ 
The (ordinary) generating function $G(z)$ for the sequence $h(n)$ is given by 
$$G(z)= \frac {10+z+z^2}{(1-z)(1-z-z^2)}.$$ 
It then follows that 
$$h(n)=2F_n+3F_{n+2}-9=2L_{n+1}+F_{n+2}-9.$$ 
\noindent 
\textbf{Definition 3.} A binomial tree $B_k$ of order $k$ ($k \geq 0)$ is an ordered tree defined recursively as follows \\
(i)  $B_0$ is a one-vertex graph.\\
(ii) $B_k$ consists of two copies of $B_{k-1}$ such that the root of one is the leftmost child of the root of the other.\\
Let $b(k)$ denote the number of D$2$CS in a  binomial tree $B_k$. Then we have
$$b(k)=2b(k-1)+2^k-2, \quad k \geq 1,  $$ 
with the initial condition $b(0)=2$.\\
By solving the above recurrence, we obtain $b(k)=k2^k+2$. Thus, $b(k)$ grows exponentially in the order of binomial tree. \\
\textbf{Definition 4.} A graph is a split if there is a partition of its vertex set into a clique and an independent set. 
\newtheorem{pro2}[pro1]{Proposition}  
\begin{pro2}
Let $G$ be a split graph with $K \subseteq V(G),|K|=\omega(G)=k$, for all $v \in K, d(v)=k+r-1$ and for all $v' \in V(G) \setminus K, d(v')=1$. Then the number of D$2$CS of $G$ is $k2^{k-1}(2^r-1)+2^k+kr$. 
\end{pro2}
\begin{proof}
From the given conditions it is clear that if $S$ is a D$2$CS of $G$ then there exists a vertex $v \in K$ such that  and $S \subseteq N[v]$. We know that the number of D$2$CS of $G$ of size $0,1$ and $2$ are respectively  $1,|V(G)|=k(r+1)$ and $E(G)=\binom{k}{2}+kr$. Now we count the number of D$2$CS of size at least $3$. Let $S$ be a D$2$CS of $G$ with $|S| \geq 3$, then $S$ fits into one of the following three cases \\
Case (i)\;\; : $|S \setminus K|=0$. Clearly the number of D$2$CS of this form is $2^k-\binom{k}{2}-k-1$. \\ 
Case (ii)\; : $|S \setminus K|=1$. The number of D$2$CS of this form is $(2^{k-1}-1)kr$.\\ 
Case (iii)\;: $|S \setminus K| \geq 2$. The number of D$2$CS of this form is $k2^{k-1}(2^r-r-1)$. \\
Therefore the number of D$2$CS of cardinality greater than two is $2^k-\binom{k}{2}-k-1+(2^{k-1}-1)kr+k2^{k-1}(2^r-r-1)$. So the total number of D$2$CS of $G$ is $1+k(r+1)+\binom{k}{2}+kr+2^k-\binom{k}{2}-k-1+(2^{k-1}-1)kr+k2^{k-1}(2^r-r-1)$, which gives the result.
\end{proof}

\section{Algorithm for Counting and Enumerating the D$2$CS of a Graph}
\noindent In this section we describe an algorithm for counting and enumerating the D$2$CS of a graph. The basic idea  is obtaining $G^2$ and generating all the cliques in $G^2$. Then all those cliques of $G^2$ which are not D$2$CS of $G$ are eliminated.
\newtheorem{fact1}{Fact}  
\begin{fact1}
Every D$2$CS in $G$ is a clique in $G^2$.
\end{fact1} 
\begin{proof}
Proof follows from the definitions of D$2$CS and $G^2$.
\end{proof} 
\noindent Algorithm \textbf{EnumAllD2CS} enumerates and counts D$2$CS of a graph $G$ with $|V(G)|=n$ and $|E(G)|=m$. The algorithm outputs the number of D$2$CS in $G$.
\begin{quote}
\textbf{Algorithm EnumAllD$2$CS(G)}
\begin{quote}
%
$1$. Enumerate all cliques in $G^2$; let
 $T_S = \{S : S \; \text{is a clique in} \; G^2 \; \& \; |S| \geq 3\}$ \\
$2$. Eliminate those elements of $T_S$ , which are not D$2$CS of $G$; \\
\hspace{10 mm} let $T_S' = \{S : S \; \text{is a D$2$CS in} \; G \; \text{and} \; |S| \geq 3\}$ \\
$3$. Return $|T_S'|+n+m+1$. 
\end{quote}
\end{quote}
\noindent \textsf{Correctness and complexity}: From Fact $1$ it is clear that \textbf{EnumAllD2CS} doesn't miss any D$2$CS of $G$. Step $2$ of the algorithm ensures that no wrong D$2$CS is generated. Hence the correctness of \textbf{EnumAllD2CS}. \smallskip \\
The complexity of \textbf{EnumAllD2CS} corresponds to the question of determining the number of D$2$CS of $G$. Clearly step $1$ of the algorithm takes O($f(n)+g(n)$) time, where $f(n)$ is the complexity of boolean matrix multiplication and $g(n)$ is the complexity of generating all the cliques in a graph on $n$ vertices and step $2$ takes O($ n^3 /log ^ 2 n$) ~\cite{Basch} time for each element of $T_S$. Summarizing, we have: 
\newtheorem{thm1}{Theorem}
\begin{thm1} 
Let $G$ be a graph. All D$2$CS in $G$ can be enumerated in O$( n^3/log ^ 2 n)$  time for each. 
\end{thm1}
\noindent \textbf{Remark.} It is easy to see that the complexity of the above algorithm is no more than O($f(n)+g(n)n^3 /log ^ 2 n$), where $f(n)$ and $g(n)$ at present stands at O($n^{2.376}$) ~\cite{Basch} and O($n^{\log_2 {n} + 2}$) ~\cite{Kresher} respectively.
\section{Linear-Time Algorithm for Enumerating Maximal D$2$CS in a Strongly Chordal Graph}
Given a graph $G$, a vertex $v \in V(G)$ is called \textit{simplicial} in $G$ if $N[v]$ induces a clique. Fulkerson and Gross ~\cite{Fulkerson} showed that a graph $G$ is \textit{chordal} if and only if it is possible to order the vertices $\{v_1,\ldots,v_n \}$ of $V(G)$ in such a way that for each $i \in $ \{$1,\ldots,n $ \}, the vertex $v_i$ is simplicial vertex of $G_i=G$ [\{$v_i,\ldots,v_n$\}]. Such an ordering is called a \textit{perfect elimination ordering} ( also referred as the \textit{p.e.o}). Let $N_i[v]$ denote the closed neighborhood of $v$ in $G_i$. The ordering of  the vertices $v_1,\ldots,v_n$ is called a \textit{strong elimination ordering} ( also referred as the \textit{s.e.o}), if it is a p.e.o and for each $i<j<k$, if $v_j,v_k \in N_i[v_i]$ then $N_i[v_j] \subseteq N_i[v_k]$. Throughout this section we assume that the vertices are numbered in s.e.o. order. \medskip \\
\textbf{Definition 5.} ~\cite{Farber} A graph $G$ is \emph{strongly chordal} if and only if it admits a s.e.o. 
\newtheorem{lem1}{Lemma} 
\begin{lem1} 
Let $G$ be a strongly chordal graph. Every maximal D$2$CS in $G$ is of the form $N[v]$, where $v \in V(G)$. 
\end{lem1}
\begin{proof}
Let $A$ be a maximal D$2$CS of $G$. We assume that $w$ and $z$ represent respectively the lowest and the highest numbered vertices of $A$. 
Now, we have three cases : \\
Case (i)\;\; : $A = N[z]$. The lemma holds.\\
Case (ii)\; : $A \subset N[z]$. It is clear that $diam$($N[z]$) $\leq 2$ and so $N[z]$ is also a D$2$CS. Hence our assumption that $A$ is a maximal D$2$CS is not correct. \\
Case (iii)\;: $A \supset N[z]$. There exist two vertices $u$ and $v$ such that $u \in A \setminus N[z]\;,v \in N[z]$ and $d(u,v) > 2$ in the graph $G[A]$; a contradiction that $A$ is a D$2$CS.\\
Hence the lemma.
\end{proof} 
\noindent The following result follows directly from the above lemma.
\newtheorem{pro3}[pro1]{Proposition} 
\begin{pro3}
Let $G$ be a strongly chordal with $|V(G)|=n$. Let $X$ be the maximum possible number of D$2$CS in $G$. Then $X \leq n$. The equality holds iff $G$ has no edges. 
\end{pro3} 
\noindent Since the converse of the lemma $2$ does not hold, the following algorithm enumerates all maximal D$2$CS in a strongly chordal graph in linear time. 
\begin{quote}
\textsf{Algorithm} \textbf{EnumMaxD$2$CSSChordal(G)}\\
\textit{Input} : A strongly chordal graph $G$ with vertices $1, \ldots ,n$ labeled in s.e.o order.\\
\textit{Output}: All maximal D$2$CS of $G.$ 
\begin{enumerate}\addtolength{\itemsep}{-0.5\baselineskip}
\item  \textbf{for} $i \leftarrow 1$ to  $n$ 
\item \hspace {3mm} $P(i) \leftarrow S(i) \leftarrow NIL$ 
\item $u \leftarrow $ largest numbered vertex of $N(1)$ 
\item \textbf{print} $N[u]$. 
\item	\textbf{for} $i \leftarrow 2$ to $n$ 
\item \hspace {3mm} $u' \leftarrow max \{v : v \in N(i), v < i \}$ 
\item \hspace {3mm} $P(i) \leftarrow max \{v : v \in N(u'), v > u'\}$ 
\item \hspace {3mm} $S(i) \leftarrow max \{v : v \in N(i), v > i \}$  
\item \hspace{0.1mm} \textbf{if} $P(i) = NIL$ \textbf{or}  $|N[S(i)] \setminus N[P(i)]| > 0$ \textbf{then}
\item \hspace {3mm} \textbf{print} $N[S(i)]$.
\end{enumerate}
\end{quote}
\medskip Clearly the algorithm generates all maximal D$2$CS of $G$ and runs in linear time. Hence we have: 
\noindent 
\newtheorem{pro4}[pro1] {Proposition} 
\begin{pro4}
All maximal D$2$CS in a strongly chordal graph on $n$ vertices can be enumerated and counted in O($n$) time. 
\end{pro4}

\end{document}